\documentclass[rnote,traditabstract]{aa}
\usepackage{subfig,graphicx}

\usepackage{multirow}
\usepackage{lscape}
\usepackage{float}
\usepackage{stfloats}
\usepackage{enumerate}
\usepackage{fixltx2e}
\usepackage{afterpage}

\usepackage{natbib}
\bibpunct{(}{)}{;}{a}{}{,}
\restylefloat{figure}

\begin{document}

\title{Viscous timescale in high mass X-ray binaries}
\author{B. \.{I}\c{c}dem%\inst{1}
\and A. Baykal%\inst{1}} 
%\and Marat Gilfanov\inst{3,4}
}

\institute{Physics Department, Middle East Technical University, Ankara 06531, Turkey
%\and Visiting Scientist, Max-Planck-Institut f\"ur Astrophysik, 85741 Garching b. M\"unchen, Germany
%\and Max-Planck-Institut f\"ur Astrophysik, 85741 Garching b. M\"unchen, Germany
%\and Space Research Institute, Russian Academy of Sciences, Profsoyuznaya 84/32, 117997 Moscow, Russia
}

\date{Received 23 September 2010 / Accepted 5 February 2011}

\abstract
{\emph{Context.}
Low mass X-ray binaries were found to have very low frequency breaks in their power
density spectra
 below which the power density spectra are nearly in white noise structure and
 at higher frequencies they approximately follow the $P_\nu \propto \nu^{-1.3}$ law.

\emph{Aims.} 
%In 2005, Gilfanov and Arefiev studied X-ray variability of persistent LMXBs
% in the $10^{-8}-10^{-1}$ Hz frequency range and 
 To determine whether high mass X-ray binary power density spectra
have similar properties and the findings for low mass X-ray binaries are 
also valid for high mass binaries, we analyzed the time series of high mass X-ray
 binary sources produced by All Sky Monitor of Rossi X-ray Timing Explorer.

\emph{Method.} 
 We obtained the power density spectra of the high mass X-ray binaries using
 the cosine transform of autocorrelation function.

\emph{Results.}
We identified break frequencies 
for  seven sources, namely OAO 1657-415, SS 433, Vela X-1, 
SMC X-1, 4U 1700-377, GX 301-2, and LMC X-1.
 The normalized break frequencies with respect to the orbital frequency
 ($f_{break}/f_{orbit}$) for sources
OAO 1657-415, SS 433, SMC X-1 and LMC X-1
 are consistent with those of Roche lobe overflow systems. 
The other high mass X-ray binary systems, 
Vela X-1, GX 301-2, and 4U 1700-377, however, 
have larger  break frequency ratios, $f_{break}/f_{orb} $, 
which are indicative of short viscous times.
 These are all wind-accreting sources and the stellar winds
 in the systems allow the formation of only
 short radius discs.
 Consequently, we qualitatively distinguished the Roche lobe overflow binaries from the
wind accreting system by comparing their normalized break frequencies.
}


\maketitle

\section{INTRODUCTION}

X-ray binaries have X-ray flux variations on a broad range of timescales.
 In a binary system the transfer of matter through an accretion disc is considered to be dominated by viscous forces.
 These modify the disc structure on a viscous timescale.
 Hence, the response of the disc to any external change occurs on this longest timescale of the disc.
 If the frequency of a variation is lower than $t_{visc}(R_d)^{-1} \sim f_{break}$,
 the corresponding X-ray flux
 variations are considered to be uncorrelated; therefore, the corresponding power density spectrum is
 expected to exhibit a white noise behavior at these frequencies \citep{gil05}. 

\citet{gil05} found that the power density spectra of some persistent low mass X-ray binary systems
 have characteristic shape: white noise up to a certain frequency value after which a power law of
  $P \propto \nu^{-\alpha}$ appears where $\alpha \sim1.3$. 
%In order to determine whether their results are compatible
% with the theoretical viscous time scale predictions, they compared these with the break frequencies of the
% power spectra they determined. 
%They found that the power spectra of 11 LMXRB sources had a very low frequency 
%break and discovered that the break frequency correlated with the orbital frequency.
%However, the measured values of the break frequency imply about an order of magnitude smaller viscous 
%time than that expected by the standard accretion disc theory \citep{sha73} 

In this study, we find significant break frequencies  
in OAO 1657-415, SS 433, Vela X-1, SMC X-1, 4U 1700-377, GX 301-2, and LMC X-1. 
In the following section, we explain the selection criteria and our sample of X-ray binaries.
 We also the describe the techniques we used to obtain the power density spectra,
 and then determine the break frequencies together with the power law indices.
Discussion of our results is given in section 3.

\section{DATA \& ANALYSIS}
\subsection{Data}
We used data gathered by the All Sky Monitor (ASM) instrument of Rossi X-ray Timing Explorer (RXTE) observatory,
 covering the period from 1996 to 2010. The ASM instrument operates in the 2-12 keV energy range,
 performs flux measurements for over $\sim500$ X-ray sources, and scans 80\% of the sky every 90 minutes.
 Each flux measurement has a duration of $\sim90$ seconds.
 Some gaps sometimes appear in time as long as a few months in the light curves of
 ASM sources because of navigational constrains and the appearance of very bright transient sources.
 The dwell-by-dwell light curves were retrieved from the public RXTE/ASM archive at HEASARC.

We derived the power density spectra (PDS) of the sources from the cosine transform
 of the discrete autocorrelation function.  
This technique has advantages if the time series are unevenly sampled 
\citep{gil05}.
 We also estimated the PDS using HEASARC software "powspec" with 
Miyamoto normalization \citep{miy92} in our data analysis.
Both estimations of PDS give consistent results.

 \subsection{Sources}
We constructed our sample after evaluating the PDS 
of light curves for 45 high mass X-ray binaries with known orbital periods.

\begin{figure*}[ht!]
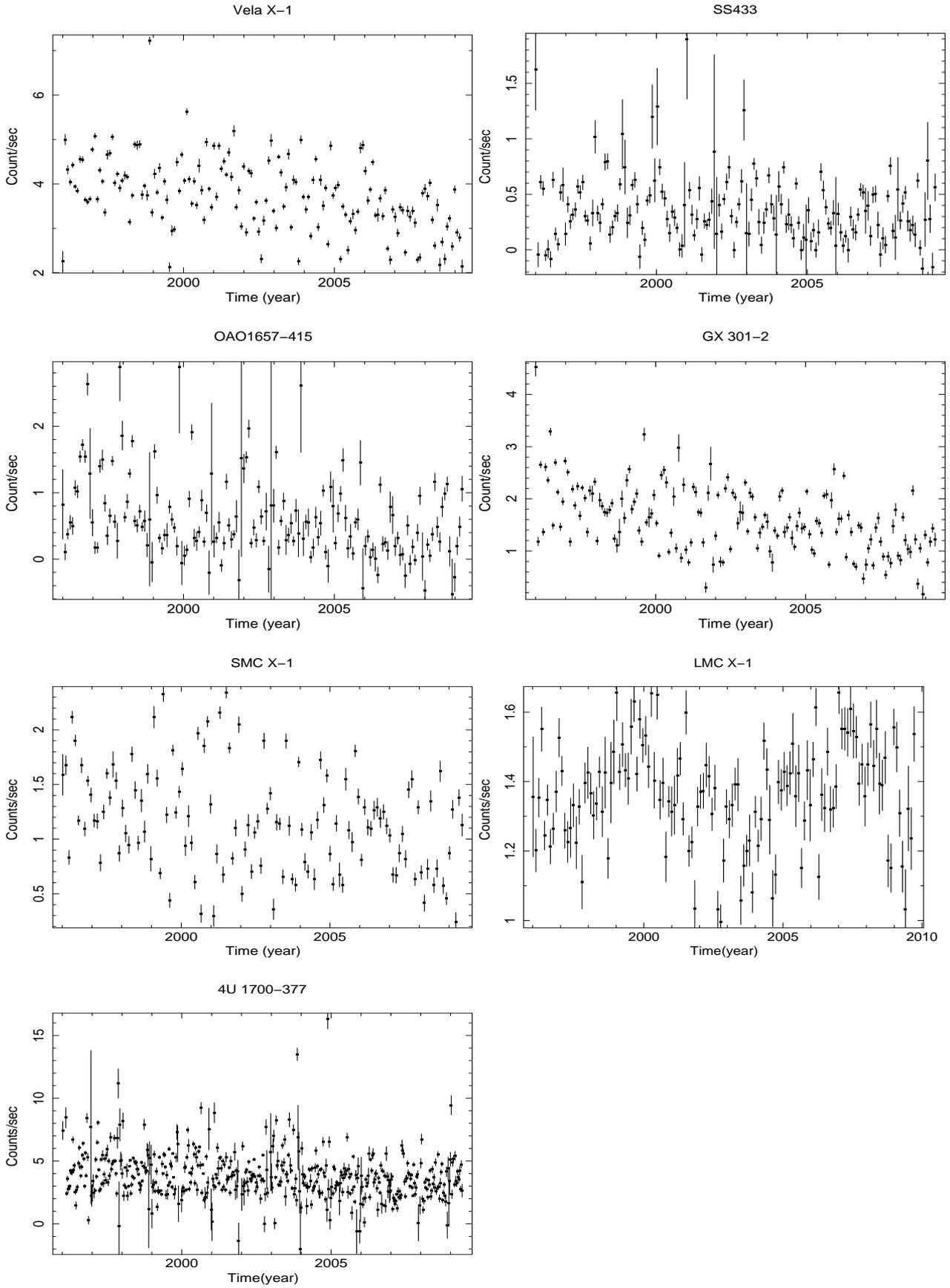

\subfloat{
\includegraphics[angle=-90,width=0.45\textwidth,totalheight=0.22\textheight]{velax-1_lc_year}} 
\subfloat{
\includegraphics[angle=-90,width=0.45\textwidth,totalheight=0.22\textheight]{ss433_lc_year}} \\
\subfloat{
\includegraphics[angle=-90,width=0.45\textwidth,totalheight=0.22\textheight]{oao1657-415_lc_year}} 
\subfloat{
\includegraphics[angle=-90,width=0.45\textwidth,totalheight=0.22\textheight]{gx301-2_lc_year}} \\
\subfloat{
\includegraphics[angle=-90,width=0.45\textwidth,totalheight=0.22\textheight]{smcx-1_lc_year}} 
\subfloat{
\includegraphics[angle=-90,width=0.45\textwidth,totalheight=0.22\textheight]{lmcx-1_lc_year}} \\
\subfloat{
\includegraphics[angle=-90,width=0.45\textwidth,totalheight=0.22\textheight]{x1700-377_lc_year}} 
\caption{\small{Long-term light curves of the high-mass X-ray binaries}}
\label{fig:light_curves}
\end{figure*}

\begin{table*}
\begin{center}

\caption{The binary parameters for LMXBs from the ASM sample}
\label{tab:source_parameters}
\begin{tabular}[htb]{ccccccc}
\hline
\hline
\multirow{2}{*}{Source} &
\multirow{2}{*}{$M_1(M_{\sun})$} &
\multirow{2}{*}{$M_2(M_{\sun})$} & \multirow{2}{*}{$P_{orb}(days)$} & %\multirow{2}{*}{$a(10^{10}$cm)} & 
\multirow{2}{*}{$q$} & \multirow{2}{*}{$D$(kpc)} & \multirow{2}{*}{$Ref.$} \\ & & & & & & \\
\hline
OAO1657-415 & $1.4$ & 14-18 & $10.4436$ & %\pm0.0038$ & %$106.0/\sin(i)$ & 
10-12.9 & $6.4\pm1.2$ $^8$ & 1 \\
%\hline
SS433 & $\sim9$ & $\sim30$ & 13.082 & %740 & 
$\sim3.33$ & 5 & 2 \\
%\hline
\multirow{2}{*}{Vela X-1} & $2.27\pm0.17$ & $27.9\pm1.3$ & \multirow{2}{*}{$8.964368$}&%\pm0.000040$} & %410.4 & 
$12.29\pm0.35$ & \multirow{2}{*}{$1.9\pm0.2$ $^{3}$} & \multirow{2}{*}{4} \\
& $1.88\pm0.13$ & $23.1\pm0.2$ & & %341.9 & 
$12.29\pm0.74$ & & \\
%\hline
GX 301-2 & $1.85\pm0.6$ & 39-53 &
$41.5$ &  
 $21.74\pm6.62$ & $3-4$ & $5$ \\
%\hline
SMC X-1 & $1.06\pm0.1$ & $17.2\pm0.6$ &
$3.8923$ &  
 $16.23\pm2.10$ & $61$ & $6$ \\
%\hline
LMC X-1 & $10.91\pm1.41$ & $31.79\pm3.48$ &
$3.90917$ &%\pm0.00005$ & 
 $2.91\pm0.70$ & $\sim50$ & $7$ \\
%\hline
4U 1700-377 & $2.44\pm0.27$ & $58\pm11$ &
$3.412$ &  
 $23.77\pm7.14$ & $1.8$ & $8$ \\
\hline
\end{tabular}\\
\end{center}

 {\bf Notes.} $M_1$ and $M_2$ are masses of the
primary and secondary companions, respectively. $P_{orb}$ is the orbital period, $a$ is the binary separation, $q$
is the mass ratio $M_1/M_2$, and $D$ is the distance of the system. The mass of the pulsar of OAO 1657-415 has not been determined to be a certain value and we assumed it to be $1.4M_{\sun}$ in our calculations. The two sets of parameters for
HMXRB corresponds to inclination angles $i=70.1\deg$ and $90\deg$, respectively. $Ref.$ refers to References for the parameters:%$^1$ \citet{har04};$^2$ \citet{fen99};$^3$ \citet{ban99};$^4$ \citet{john99};$^5$ \citet{oro99};$^6$ \citet{wat96};$^7$ \citet{ste02};$^8$ \citet{chak02};
(1) \citet{chak93}; (2) \citet{cher05}; (3) \citet{sad85}; (4) \citet{qua03}; (5) \citet{kap06}; (6) \citet{meer07}; (7) \citet{oro09}; (8) \citet{cla02}%; $^{17}$ \citet{del01}; $^{18}$ \citet{sor01}

\end{table*}

\begin{figure*}[ht!]
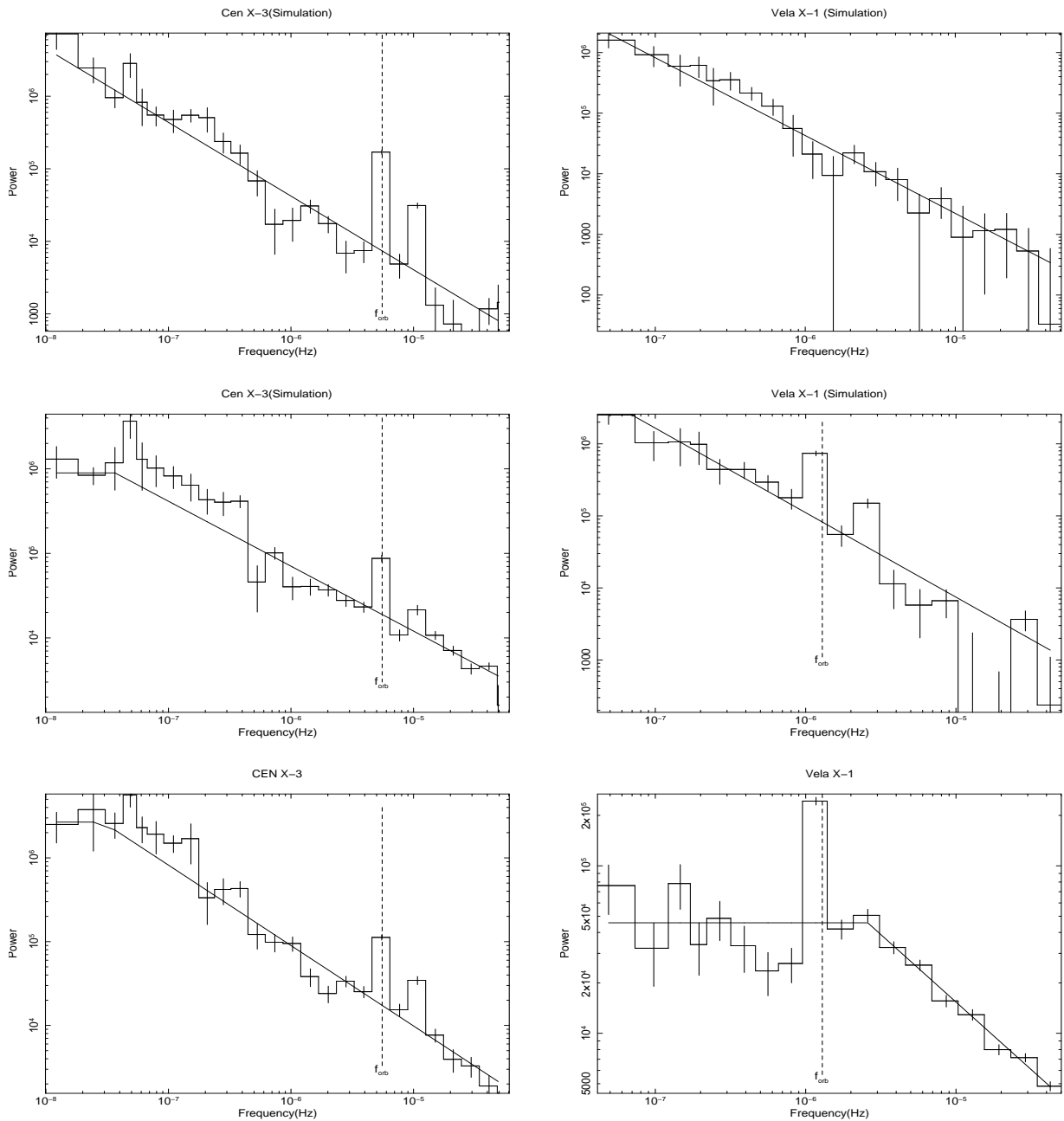

\subfloat{
\includegraphics[trim=0cm 0cm 0cm 1cm, clip=true, angle=-90,width=0.45\textwidth,totalheight=0.22\textheight]{fake_cenx3_orbmod_nobreak_notrans}}
\subfloat{
\includegraphics[trim=0cm 0cm 0cm 1cm, clip=true, angle=-90,width=0.45\textwidth,totalheight=0.22\textheight]{fake_velax1_noorbmod_nobreak}} \\
\subfloat{
\includegraphics[trim=0cm 0cm 0cm 1cm, clip=true, angle=-90,width=0.45\textwidth,totalheight=0.22\textheight]{fake_cenx3_orbmod_nobreak}}
\subfloat{
\includegraphics[trim=0cm 0cm 0cm 1cm, clip=true, angle=-90,width=0.45\textwidth,totalheight=0.22\textheight]{fake_velax1_orbmod_nobreak}} \\
\subfloat{
\includegraphics[trim=0cm 0cm 0cm 1cm, clip=true, angle=-90,width=0.45\textwidth,totalheight=0.22\textheight]{cenx-3_lowfreq_pds}} 
\subfloat{
\includegraphics[trim=0cm 0cm 0cm 1cm, clip=true, angle=-90,width=0.45\textwidth,totalheight=0.22\textheight]{velax-1_lowfreq_pds_2048}}
\caption{\small{On the left side, the simulation results of Cen X-3 is presented. The top figure is the power density spectrum
 in power law with orbital modulation, the middle one is the same as the former but the state transitions 
are taken into account, and the bottom figure is the real PDS of Cen X-3. On the right side, the simulation results of 
Vela X-1 is presented. The top figure is the power spectrum
only in power law, the middle one is the PDS in power law with orbital modulation, and the bottom figure is the real 
PDS of Vela X-1. It is clear from the figures that sinusoidal orbital modulation does not affect the trend of the power
law, but the state transitions cause an artificial break in the PDS.}}
\label{fig:Simulations}
\end{figure*}

\begin{figure*}[ht!]
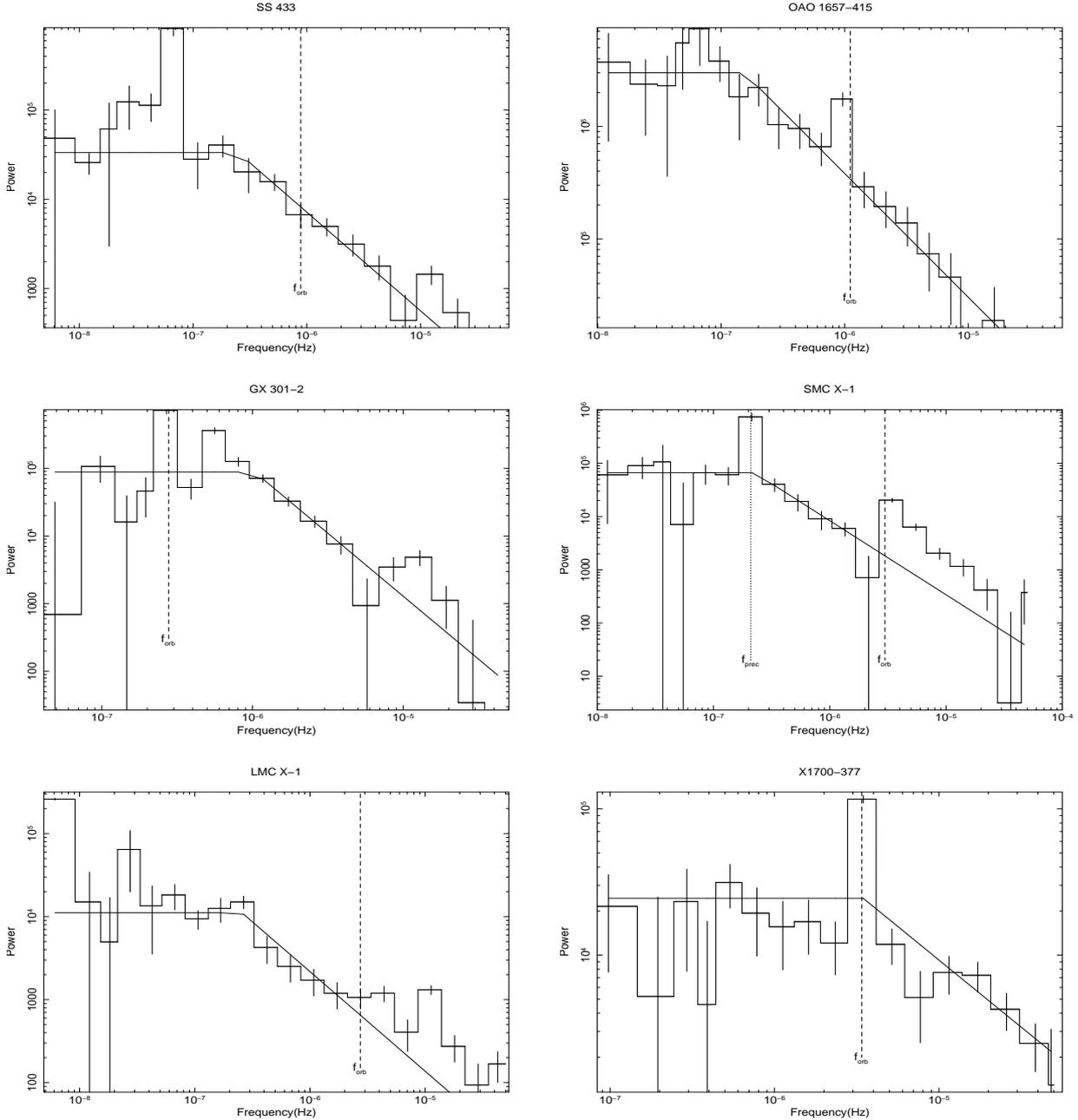

%\subfloat{
%\includegraphics[trim=0cm 0cm 0cm 1cm, clip=true, angle=-90,width=0.45\textwidth,totalheight=0.22\textheight]{velax-1_lowfreq_pds_2048}}
\subfloat{
\includegraphics[trim=0cm 0cm 0cm 1cm, clip=true, angle=-90,width=0.45\textwidth,totalheight=0.22\textheight]{ss433_lowfreq_pds_16394}} 
\subfloat{
\includegraphics[trim=0cm 0cm 0cm 1cm, clip=true, angle=-90,width=0.45\textwidth,totalheight=0.22\textheight]{xa_x1657-415_d1_asm_test_max_like_fit}} \\
\subfloat{
\includegraphics[trim=0cm 0cm 0cm 1cm, clip=true, angle=-90,width=0.45\textwidth,totalheight=0.22\textheight]{gx301-2_lowfreq_pds_2048}} 
\subfloat{
\includegraphics[trim=0cm 0cm 0cm 1cm, clip=true, angle=-90,width=0.45\textwidth,totalheight=0.22\textheight]{smcx-1_lowfreq_pds_8192}} \\
\subfloat{
\includegraphics[trim=0cm 0cm 0cm 1cm, clip=true, angle=-90,width=0.45\textwidth,totalheight=0.22\textheight]{lmcx-1_lowfreq_pds_16394}} 
\subfloat{
\includegraphics[trim=0cm 0cm 0cm 1cm, clip=true, angle=-90,width=0.45\textwidth,totalheight=0.22\textheight]{x1700-377_lowfreq_pds_1024}} 
%\subfloat{
%\includegraphics[trim=0cm 0cm 0cm 1cm, clip=true, angle=-90,width=0.5\textwidth,totalheight=0.2\textheight]{xper_nooutburst_lowfreq_pds_8192}} \\
%\subfloat{
%\includegraphics[trim=0cm 0cm 0cm 1cm, clip=true, angle=-90,width=0.5\textwidth,totalheight=0.2\textheight]{lmcx-3_lowfreq_pds_16384}} 
\caption{\small{Power density spectra of high mass X-ray binaries with the best power-law models 
(The PDS of Vela X-1 is presented in Fig.\ref{fig:Simulations} at the bottom right corner)}}
\label{fig:power_spectra}
\end{figure*}

\begin{enumerate}[(i)] 
\item  We selected only persistent sources that are in the same state during the observation period.
  Twenty-five HMXRBs in our sample are Be transients that are not persistent X-ray
 sources and display outbursts because of the extremely wide eccentric orbit of neutron stars around their companions.
Owing to the short outburst time span relative to the orbital period, we do not see breaks in their PDS. 
 Outburst phases are the cause of substantial red noise in their PDS.
 Two sources Cen X-3 and Cyg X-1 have low frequency breaks in their PDS.
 However, these sources have significant signatures of high-low states in their light curves.
  These state transitions contaminate the long-term power density spectra, so that 
 fake breaks may appear. To see these artificial effects, we carried out simulations
as described below in (iii). In addition, we refer to Fig.
 \ref{fig:Simulations} for the comparison between the PDS obtained from the real light curve
  and the simulated light curve. While creating the fake light curve, we took into account
  the state transitions of the source. Although we produced this light curve from a PDS obeying 
  a power law trend (i.e. no breaks), it is clear from the figure that the simulated PDS
  resembles the real PDS. Hence,
 we estimate that the breaks in these sources are associated with 
 the occurrence of high-low intensity states rather than the viscous time of the disc. 

\item The sources with very low count rates, $<1$ count/sec, such as LMC X-4 or 4U1538-52, 
have white-noise-dominated power density spectra at low frequencies.

 \item In our analysis, we found that the remaining seven sources  
 % LMC X-3 and 4U 0352+309
  had low frequency breaks in their PDS together with 
 power-law indices  $\sim1.3$ 
consistent with the LMXRBs of \citet{gil05}. These are 
OAO 1657-415, SS 433, Vela X-1, SMC X-1, 4U 1700-377, GX 301-2, and LMC X-1.
These values for $\alpha$ indices imply the existence of accretion discs \citep{lyub97}.
 In Figs. \ref{fig:light_curves} and
  \ref{fig:power_spectra} and Table \ref{tab:source_parameters} 
light curves, PDS and orbital parameters of the sources are presented,
 respectively. We marked the orbital and super-orbital periods of the
sources on the power density spectra. Although these modulations are quite
clear as peaks at the corresponding frequencies, it is not possible to
determine their values from the PDS accurately since we use geometrical binning and 
the resolutions of the PDS are low.
Table 2 contains the break frequencies and power-law indices of the sources. 

 Power spectra of light curves of the sources show that there are periodic modulations in the
 luminosity of the sources that are caused by both the orbital motion of the system and  
the quasi-periodic/periodic modulation due to the precession of accretion disc.
To see whether the low frequency breaks, which we saw in the PDS of the HMXBs,
are the results of such periodic/quasi-periodic variability in the X-ray flux,
 we carried out
simulations using the orbital period and precession periods of each source. 
In simulating the light curves of each source,
we took Fourier amplitudes from the real power spectra, and
for the phases of Fourier amplitudes, we
used Gaussian white noise with unit variance.
We sampled the light curves of each source as 
if observed by ASM such that their power densities are 
consistent with the power law 
$P\sim{f^{-1.3}}$. 
 We then convolved these  time series by introducing 
periodic signals for both the orbital motion and precession period if the source had a periodicity.
Consequently, using the simulated light curves
 we constructed the power density spectrum.
In our simulations, we did not see any breaks due to the orbital or 
precession modulation of light curves. 
The simulation result for Vela X-1 is presented in Fig.\ref{fig:Simulations}.
 
 Other short-term variations, such as flickering and magnetic field effects
on the inner disc, appear at high frequencies, i.e. higher than our 
Nyquist frequency $>1.0\times10^{-4}$ Hz, so they do not 
'pollute' the low frequency power density spectra we analyze.  
These short-term variations are the subject of independent research. 
(For magnetic effects, see \citet{rev09}.)
Therefore, all of the seven sources were chosen for further analysis in this work.

\end{enumerate}

\section{DISCUSSION}

 The primary goals of this work are to study the low-resolution PDS of HMXBs and both resolve 
low frequency breaks  and distinguish them on the basis of their accretion properties.
 According to the standard theory of the alpha

\begin{table*}
\caption{The parameters of power spectrum approximation}
\begin{tabular}[b]{ccccccc}
\hline
\hline
\multirow{2}{*}{Source} &
\multirow{2}{*}{(count/sec)} &
{$L_x$} & \multirow{2}{*}{$f_{break}$(Hz)} & 
\multirow{2}{*}{$\alpha$} & \multirow{2}{*}{$f_{range}$(Hz)} & 
\multirow{2}{*}{$f_{break}/f_{orb}$} %& \multicolumn{2}{|c}{\multirow{3}{*}{\citet{gil05}'s Results}} 
\\
& & $(10^{37}$ergs/s) & & & & \\ 
\hline
\\

{OAO1657-415} & {2.25} & {0.3} & 
{$1.74^{+0.30}_{-0.30}\times10^{-7}$} & {$1.15^{+0.12}_{-0.09}$} & 
{$2.0\times10^{-8}-4.0\times10^{-5}$} & {$0.157^{+0.027}_{-0.027}$} 
\\
%\hline
{SS 433} & {1.53} & {0.1} & 
{$4.59^{+0.83}_{-0.83}\times10^{-8}$} & {$1.22^{+0.11}_{-0.09}$} & 
{$2.0\times10^{-8}-4.0\times10^{-5}$} & {$0.052^{0.009}_{-0.009}$}
%& \multirow{2}{*}{--} & \multirow{2}{*}{--} \\  
%& & & & & & 
\\
%\hline
{Vela X-1} & {5.00} & {0.4} & 
{$3.30^{+0.18}_{-0.20}\times10^{-6}$} & {$0.81^{+0.03}_{-0.03}$} & 
{$2.0\times10^{-8}-4.0\times10^{-5}$} & {$2.556^{+0.155}_{-0.155}$}
%& \multirow{2}{*}{--} & \multirow{2}{*}{--} \\ 
%& & & & & &  
\\
%\hline
{GX 301-2} & {1.58} & {$0.8-2.0$} & 
{$1.03^{+0.17}_{-0.16}\times10^{-6}$} & {$1.86^{+0.30}_{-0.25}$} & 
{$4.0\times10^{-8}-4.0\times10^{-5}$} & {$1.47^{+0.22}_{-0.22}$}
%& \multirow{2}{*}{--} & \multirow{2}{*}{--} \\ 
%& & & & & &  
\\
%\hline
{LMC X-1} & {1.36} & {$\sim23$} & 
{$2.57^{+0.42}_{-1.01}\times10^{-7}$} & {$1.20^{+0.31}_{-0.31}$} & 
{$5.0\times10^{-9}-4.0\times10^{-5}$} & {$0.087^{+0.015}_{-0.034}$} 
%& \multirow{2}{*}{--} & \multirow{2}{*}{--} \\  
%& & & & & & 
\\
%\hline
%\multirow{2}{*}{LMC X-3} & \multirow{2}{*}{1.34} & \multirow{2}{*}{$\sim20$} & 
%\multirow{2}{*}{$6.90^{+1.78}_{-1.59}\times10^{-8}$} & \multirow{2}{*}{$2.07^{+0.81}_{-0.48}$} & 
%\multirow{2}{*}{$5.0\times10^{-9}-4.0\times10^{-5}$} & \multirow{2}{*}{0.01} & 
%\multirow{2}{*}{--} & \multirow{2}{*}{--} \\ 
%& & & & & & & & \\
%\hline
{SMC X-1} & {1.22} & {$55$} & 
{$2.25^{+0.9}_{-0.9}\times10^{-7}$} & {$1.39^{+0.45}_{-0.45}$} & 
{$1.0\times10^{-8}-4.0\times10^{-5}$} & {$0.076^{+0.036}_{-0.024}$} 
%& \multirow{2}{*}{--} & \multirow{2}{*}{--} \\ 
%& & & & & & 
\\ 
%\hline
{4U 1700-377} & {3.83} & {$\sim0.1$} & 
{$2.58^{+0.43}_{-0.35}\times10^{-6}$} & {$0.83^{+0.16}_{-0.12}$} & 
{$8.0\times10^{-8}-4.0\times10^{-5}$} & {$0.76^{+0.13}_{-0.10}$}
%& \multirow{2}{*}{--} & \multirow{2}{*}{--} \\ 
% & & & & & & 
\\ 
&&&&&&
\\
\hline
%\enddata
\end{tabular}
{\bf Notes.} $L_x$ is the X-ray luminosity
calculated using respective count rates and source distances listed in Table 1. 
$\alpha$ is the appropriate power-law slope. 
%The last two column contain the findings of \citet{gil05} for low mass X-ray binary sources.

\label{tab:analysis_results}
\end{table*}
%\end{center}
%\end{landscape}

disc model (Gilfanov $\&$ Arefiev 2005, 
Shakura $\&$ Sunayev 1973), the viscous timescale of accretion disc (inverse break 
frequency)
 \noindent$t_{visc} = 1/f_{visc}\sim1/f_{break} $
 is proportional to the orbital period and this proportionality is given by 

\begin{equation}
\frac{f_{visc}}{f_{orb}}=\frac{2\pi\alpha}{\sqrt{1+q}}\left(\frac{H_d}{R_d}\right)^2\left(\frac{R_d}{a}\right)^{-3/2} \label{visc_orb_ratio}
\end{equation}
where $\alpha $ is the viscosity parameter and $H/R$ is the ratio of disc thickness to 
outer disc radius, $q$ is the mass ratio $M_{1}/M_{2}$, and $a$ is the binary separation.
  
\begin{figure*}[ht!]
\centering
\includegraphics[angle=-90,width=0.75\textwidth]{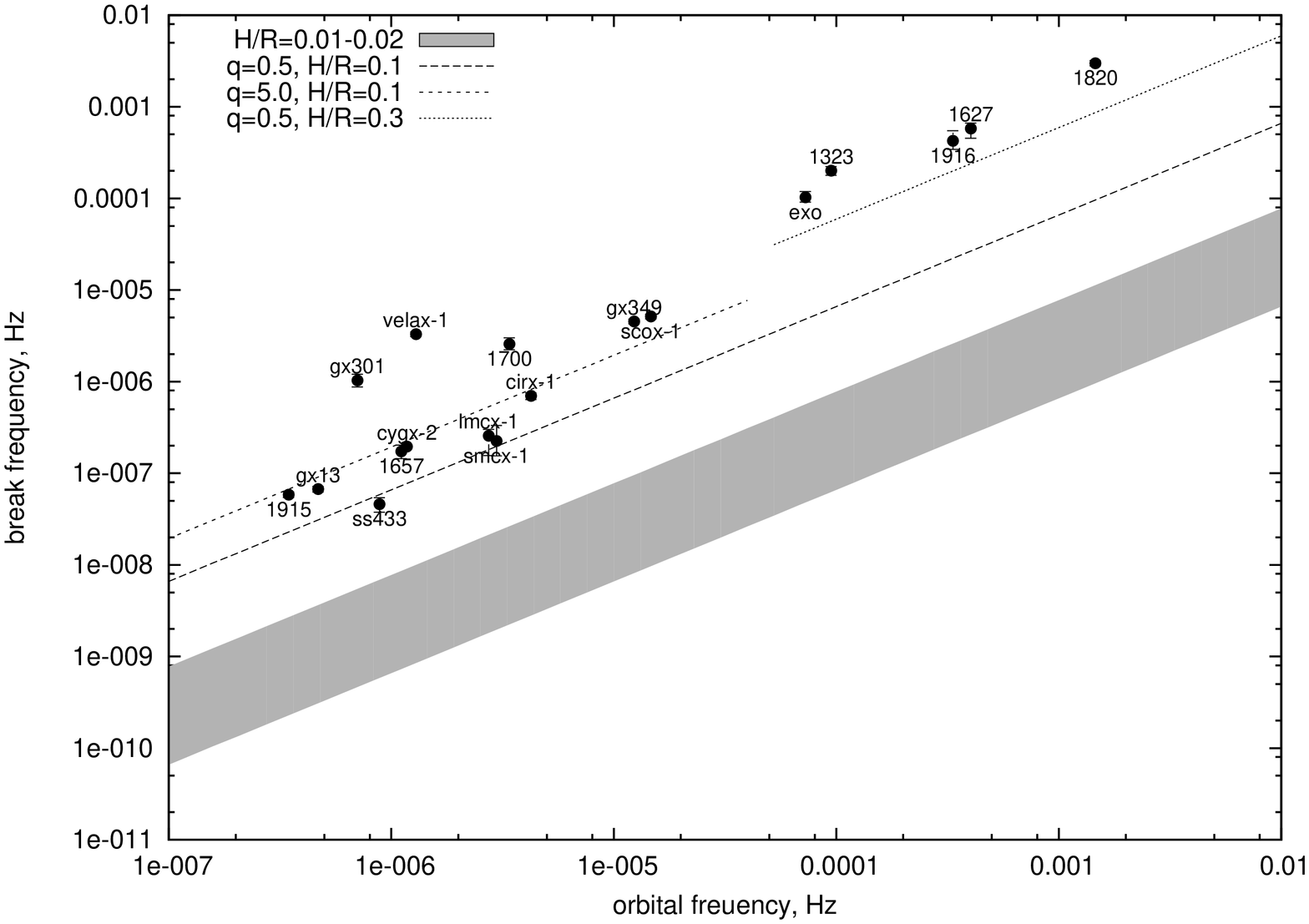}
\hspace{2cm}
%\parbox{15cm}{
\caption{\small {The relation between the PDS break frequency and the orbital frequency of the
binary system, the findings of \citet{gil05} regarding short-period LMXRBs being included.
The large shaded area toward the bottom of the plot is the widest region possible for
the dependence $f_{visc}$ vs. $f_{orb}$ according to the $\alpha$-disc, 
obtained from Eq.\ref{visc_orb_ratio} in the mass ratio range
$
2.91\leq q\leq22.77$ for $\alpha=0.5$ and
$R_d/a\sim R_{tidal}/a\sim0.112+\frac{0.270}{1+q} + \frac{0.239}{(1+q)^2}$ \citep{gil05}.
The dashed lines are predictions for larger values of the disc thickness $H/R$ with $\alpha=0.5$
and $q$ values indicated on the plot.
 Most of the errors in $f_{break}$ are on the order of marker (or symbol) size.}}%}
\label{fig:hr_curve}
\end{figure*}

In Fig. \ref{fig:hr_curve},  
 we presented the correlation between the 
 break frequency and orbital frequency of HMXBs as well as the LMXBs 
observed \citep{gil05} for various H/R and $q$ values, respectively. 

\begin{figure*}[ht]
\centering
\subfloat[]{
\includegraphics[angle=-90,trim=0cm 0cm 0cm 0cm, clip=true,width=0.48\textwidth]{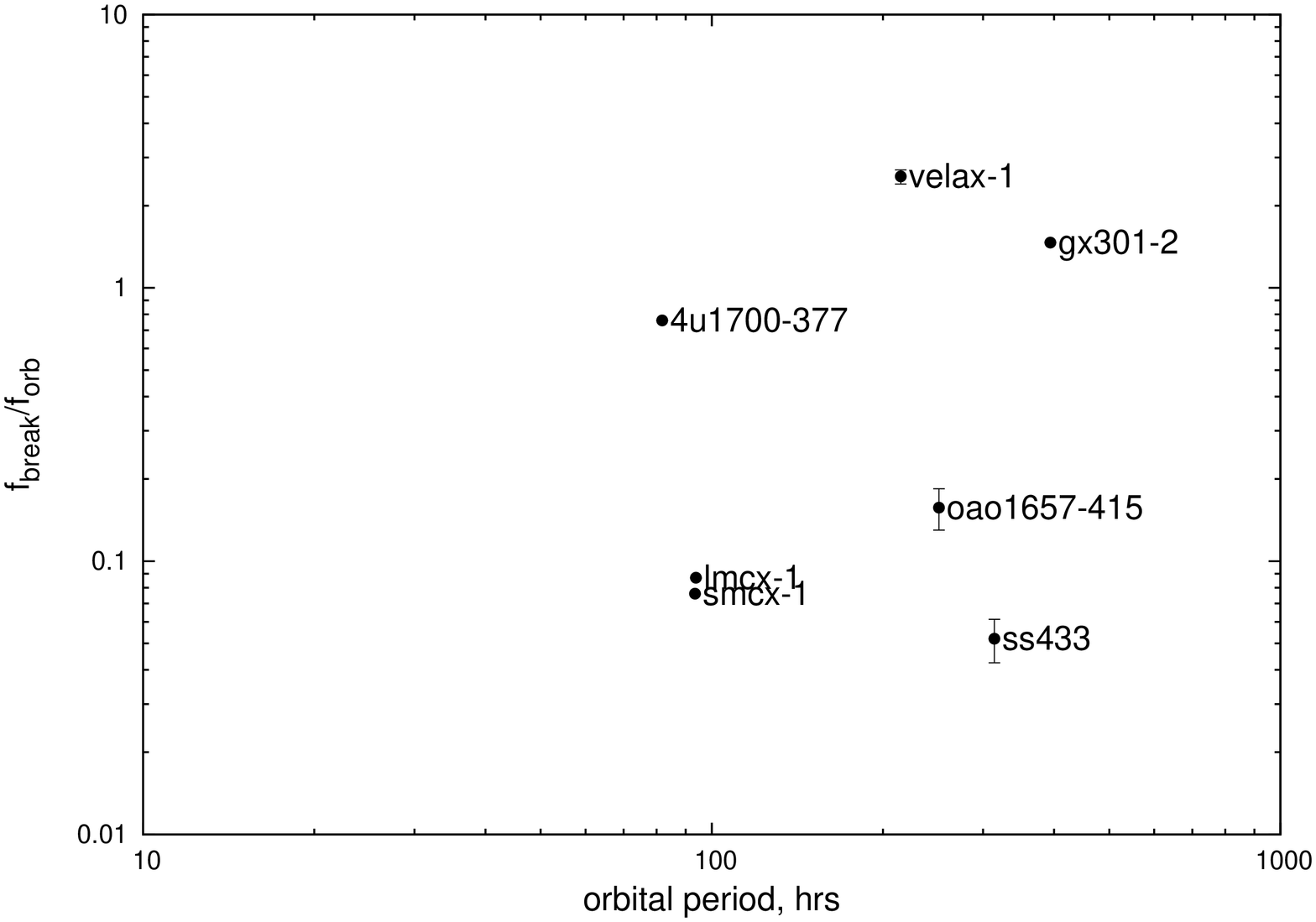}}
\subfloat[]{
\includegraphics[angle=-90,trim=0cm 0cm 0cm 0cm, clip=true,width=0.48\textwidth]{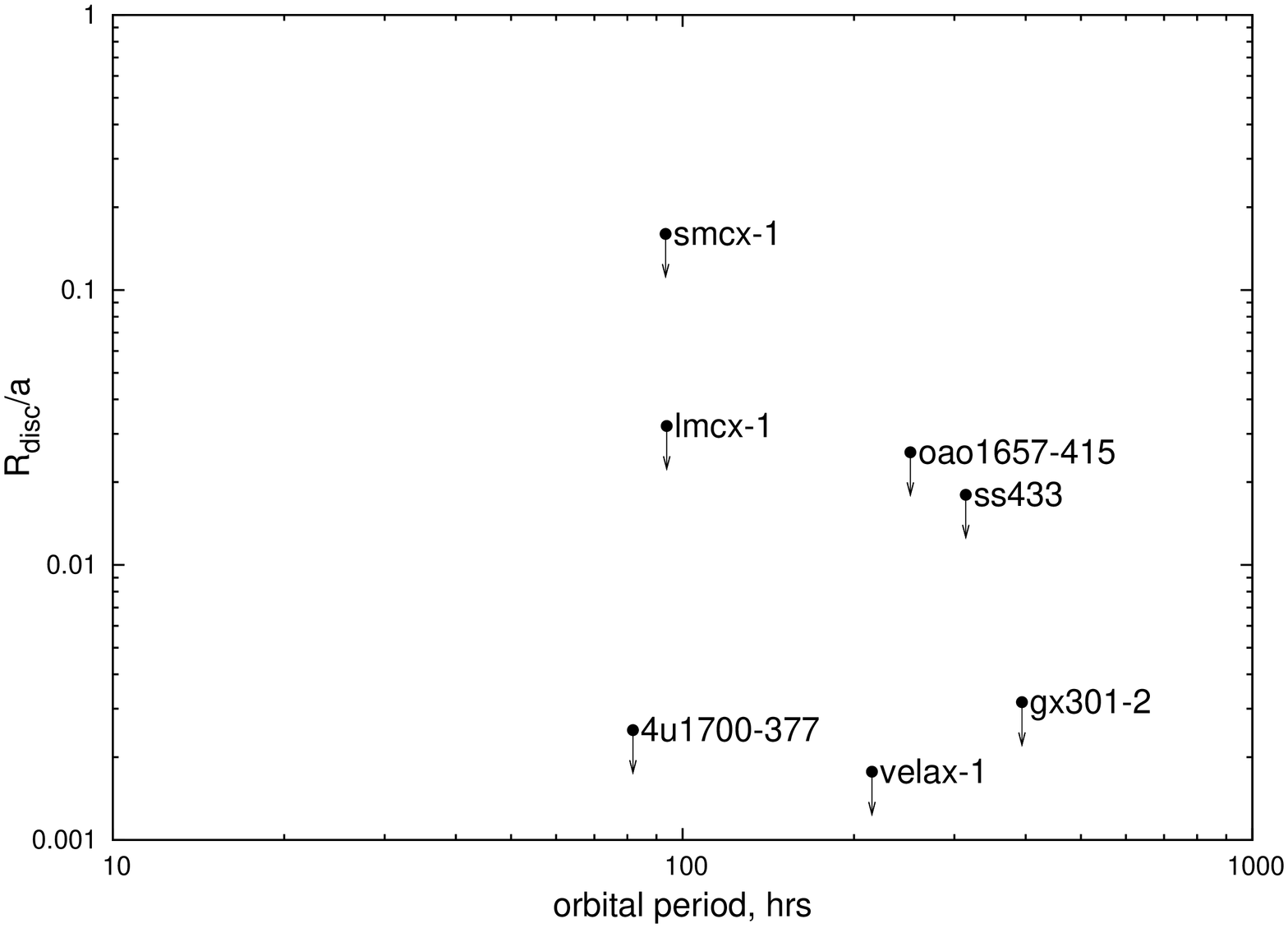}}

\caption{\small{(a)Dependence of the ratio $f_{break} /f_{orb}$ on the orbital period of the binary system.
 Some of the errors in $f_{break}/f_{orb}$ are on the order of the marker (or symbol) size.
(b)Relation of the normalized outer disc radius to the orbital period of the system.}
 The arrows represent the possible ranges of the outer disc radii \citep{blon00}.}
\label{fig:ratio_curve}
\end{figure*}

In Fig. \ref{fig:ratio_curve}-a, we  plotted the normalized 
break frequencies, $f_{break}/f_{orb}$, against the 
orbital periods.
In this figure, it can clearly be seen that, for Vela X-1, 4U 1700-377, and GX 301-2, the normalized break 
frequency is greater than for other sources (see Table 2). 
For this group of sources, normalized break frequencies are in the range 0.76-2.56, 
whereas for SS 433, SMC X-1, GX 301-2, and OAO 1657-415 these values 
are in the range 0.076-0.16. The ratios of the normalized frequencies
of the HMXBs in the former group to those in the latter group range between 5 and 33.
% This means that Vela X-1, 4U 1700-377 and GX 301-2 have 
%5-33 times higher normalized frequencies with respect to the other 
%group of sources.
 If the normalized break frequencies are associated 
with viscosity times, then Vela X-1, 4U 1700-377, and GX 301-2 should 
have tiny discs with respect to the other sources.

There should be
some reasons for this distinction regarding the properties
of accretion processes in these binaries. Since we are investigate
high mass X-ray binaries, there are two
main accretion mechanisms: Roche lobe overflow and stellar wind
accretion. Roche lobe overflow leads to persistent disc formation
around the compact object. On the other hand, thin 
accretion discs may appear in the course of wind accretion.
Among our sources, SS 433 and SMC X-1 have been
determined to be Roche-lobe overflow binaries, Vela X-1,
GX 301-2 and 4U 1700-377 involve wind-driven accretion
mechanism, and for the rest, namely OAO 1657-415 and LMC X-1, 
there are remaining uncertainties about the nature
of accretion mechanisms since they have been observed to
show signs of both Roche-lobe and wind accretion.
To understand these categories more 
clearly, we investigate
in the following section the accretion properties of 
each source.

% For systems with $f_{break}/f_{orb}>1$ 
%.in this figure it is seen that even though Vela X-1,
%4U 1700-377 
%and GX 301-2 are in wide orbital period system 
%($P_{orb} \sim 8.95$days, $3.41$days and $\sim41.5$days, respectively), considering their 
%normalized break frequencies, they are similar to the narrow orbital period systems with small $q$ values.  
% Figure \ref{fig:ratio_curve} and Table \ref{tab:analysis_results} also
% indicates that the value of $f_{break}/f_{orb}$ ranges from $\sim 0.05-0.4$
% except for those of Vela X-1, GX 301-2 and 4U 1700-377, which are 2.56, 1.47 and 0.76 respectively.

\subsection{Vela X-1}

Vela X-1 is considered to include a pulsar accreting 
from the violent wind of the supergiant companion. 
It has been observed to exhibit rapid variations 
in the periodicity of its X-ray pulses, which are indicative of spin-up and 
spin-down phases for its rotation \citep{dee89} and this 
variability is attributed to the torque reversals 
on the pulsar caused by the change in direction of the gas rotation 
 in the temporary accretion disc. This temporary disc has
been shown, by simulations, to form due to the flip-flop instability of        
the accretion shock from the stellar wind to the pulsar 
\citep{blon09}. \citet{boyn84} reported a duration of 2-3 days 
for accretion torque reversals of Vela X-1, which is close 
to the viscous time inferred from our estimate of break frequency.
Hence, the resulting disc is very unstable and cannot extend 
to a large radius in the observed 2-3 days as its dispersion 
is dominated by viscosity. 

We estimate the inner and outer radii of 
 a possible accretion disc. 
The magnetic field of the pulsar is rather 
high, $\sim10^{12}-10^{13}$ G, which that means an accretion disc is
limited by the Alfv$\acute{e}$n radius from inside given
approximately by \citep{acc02}

\begin{equation}
r_{Alfv\acute{e}n}\simeq 2.9\times10^8m_1^{1/7}R_6^{-2/7}L_{37}^{-2/7}\mu_{30}^{4/7}cm,  
\label{alfv_rad}
\end{equation}
where $m_1=M_{n}/M_{\sun}$, $R_n=R_6\times10^6$ cm, 
$L=L_{37}\times10^{37}$ erg/s, and the magnetic moment of the pulsar 
$\mu=B_nR_n^3=\mu_{30}\times10^{30}$.  
Thus, for the Vela pulsar, $r_{Alfv\acute{e}n}\simeq 3.9\times10^8$ cm. There is no
distinctive analytical or numerical expression for the outer radii 
of accretion discs of wind-driven HMXBs, but the upper limit to a possible disc radius 
is considered to be determined by the tidal forces and 
$R_{tidal}\simeq 0.9R_L$ independent of the characteristics 
of the disc ($R_L$ is the Roche-lobe radius of the compact star).
The Roche-lobe radii are calculated analytically using the expression of \citet{egg83}
\begin{equation}
 \frac{R_L}{a}=\frac{0.49}{0.6+q^{2/3}\ln(1+q^{-1/3})},
\end{equation}
which implies that $R_{tidal}\simeq 6.0\times10^{11}$ cm.
 In other words, if an accretion disc were to form,
 its radius should be greater than the Alfv$\acute{e}$n radius 
 and smaller than the tidal radius.
 This limitation is consistent with our findings since we estimate
 the outer disc radius to be $\simeq 4\times 10^{9}$ cm 
 using our value for the $f_{break}/f_{orb}$ ratio, 2.56,
 in the theoretical expression for $f_{visc}/f_{orb}$, derived from the
 Shakura-Sunyaev disc model \citep{gil05} with a thin disc 
 approximation.
 %, $H/R\sim 0.01-0.02$. 

We estimated the upper limit to the outer disc radii of the other sources with the
same procedure. Simulations carried out by \citet{blon00} estimate the disc radius 
to be smaller than our calculations. The limitation depends on many parameters such as 
mass ratio, temperature, Mach number, etc. For example, the sizes of warmer discs can be 
smaller by as much as 30\%. Nevertheless, these deviations would not change the appearance
of our results presented in the Fig.\ref{fig:ratio_curve}-b
as a plot of normalized disc radius versus orbital period.

\subsection{GX 301-2}

The mass ratio of GX 301-2 is similar to that of Vela X-1, which is 
$\sim 21.7$. It has a highly eccentric orbit,
$e\sim0.467$ \citep{sato86} with orbital period 41.5 days.
The source experienced some spin-up episodes since 1984, which are
interpreted as the signs of the formation of 
equatorial accretion discs from tidal streams found to last for
$\sim20$ days by \citet{koh97}, twice in an orbital period. 
Our analysis yielded an inverse break frequency
that corresponds to $\sim11$ days.  
 The duration of the temporary accretion disc and the timescale we obtained from 
its break frequency are of the same order implying that it has a
small accretion disc, as Fig.\ref{fig:ratio_curve}-b also indicates.

\subsection{4U 1700-377}

4U 1700-377 is an X-ray eclipsing binary system with an orbital period 
of 3.4 days \citep{jon73}. The mass of the compact object of the system
was found to be $\sim2.44M_{\sun}$ \citep{cla02}; however, it is
still unclear whether it is a neutron star or a black hole since no
X-ray periodicity has been detected that may correspond to the pulse 
period of a neutron star. Nevertheless, it is more common to accept that
it is a neutron star owing to the similarity of its X-ray spectrum to
those of accreting neutron stars, and account for the lack of pulsations 
as a result of either a weak magnetic field or an alignment of the 
magnetic field with the spin axis \citep{cla02}. 

The accretion process is considered to be driven by a wind flow rather 
than the Roche-lobe overflow because the companion star, HD 153919, 
under-fills its Roche lobe \citep{con78}. Moreover, the $\sim13.8$ day 
variations in the X-ray emission and episodic detection of QPOs appear to 
indicate the formation of a short-scale transient disc \citep{hon04}. 
Its break frequency value 
corresponds to $\sim4.4$ days.
% On the 
%other hand, \citet{hon04} has studied the long term 13.8 day variation 
%and they have originated the variation using the radiation-driven 
%warped-disc precession model. 
The normalized break frequency, 0.76, is only a factor of 2-3 
smaller then 
Vela X-1 and GX 301-2, which indicates that its accretion disc may 
be of similar size for these sources. This can also be deduced from 
its location on Fig.\ref{fig:ratio_curve}-b.

\subsection{SMC X-1}

SMC X-1 is a well-known Roche-lobe-overflow-powered X-ray binary
\citep{pap79}. It is one of the most powerful X-ray pulsars and 
is the only X-ray pulsar displaying evidence of steady spin-up \citep{woj98}.
It also has a super-orbital period of on average $\sim 55$ days,
which is assumed to be due to the precession of a warped disc \citep{woj98}.
%Precessions indicates more complicated disc structures than 
%the standard model of Shakura-Sunyaev.
 As seen in the figure, its location is consistent with 
other Roche lobe overflow systems. Its normalized break frequency is 
very low at 0.076, implying that it has a large accretion disc.

\subsection{OAO 1657-415}

OAO 1657-415 has been proven to be neither a Roche-lobe 
overflow system nor a wind-accreting system although it appears 
to contain an accretion disc \citep{bay97,bay00}. Torque reversals
have also been observed to as in the 
case of Vela X-1, but these have timescales of weeks, in contrast
to the short torque reversal period of Vela X-1 \citep{fin97}. The timescale 
of torque reversal and the inverse break frequency $\sim 66$ days are 
similar. Its normalized break frequency is 0.16, and it should 
have a large disc. On the other hand, that OAO 1657-415 has a unique place in the 
Corbet diagram in-between supergiant under-filled Roche Lobe overflow 
and Be transients suggests that there is a longer living accretion disc relative 
to the supergiant wind-fed systems.

\subsection{SS 433}

SS 433 is a black hole binary.
It has already been reported to have a break in its X-ray spectral
band at $\sim10^{-7}$ Hz by \citet{rev06}. Our finding is
 consistent with their break frequency. 
 It is known to have a Roche-lobe-overflow-powered accretion disc
similar to the low mass X-ray binaries \citep{beg06}.Thus,
it is expected to be somewhat similar to low mass X-ray
binaries with their accretion disc features. That is also
confirmed by 
its position on the plot of the correlation of break frequency
to orbital frequency (see Fig.\ref{fig:hr_curve}). 
Its normalized break frequency, $0.05$, 
is close to that of SMC X-1, which indicates that they have similar disc properties.

\subsection{LMC X-1}

LMC X-1 is a persistently luminous X-ray source in the Large
Magellanic Cloud. The estimates for the compact star of the
binary makes it a good candidate for being a black hole binary.
 It is also unusual because it is a wind driven system with a steady state disc. 
The reason why it is considered to be 
a wind accretor despite having a disc is that the optical counterpart
of the compact star fills only 45\% of its Roche lobe \citep{oro09}. Since a wind-driven
system has a disc of smaller radius than a Roche-lobe-overflow binary, 
the disc of LMC X-1 is expected to have a shorter
viscous timescale than the Roche-lobe overflow systems.
%In addition, \citet{oro09} suggest radiation reprocessing
%for the driving mechanism of the disc rather than viscous forces. If
%this is true, the time scale of the disc will be found to be much
%smaller than the viscous time scale. 
However, its normalized break frequency, 0.087, is 
very low and its location
 on the $f_{break}$-$f_{orb}$ correlation plot
is among Roche-lobe-overflow binaries despite being expected
to be near the group of Vela X-1. 
\citet{ruh10} claim that LMC X-1 is neither a Roche-lobe-overflow
binary nor a wind-accretor, but sits at the boundary. Their reasoning 
is that the emission and spectral properties of the source do not
belong to any specific group. Our result here is consistent with these
findings. A more detailed analysis should be performed to account
for the unusual behavior of this HMXB.

Figs. \ref{fig:ratio_curve}-a and \ref{fig:ratio_curve}-b clearly show the distinction 
between the Roche lobe overflow systems and the wind accreting systems. 
 The normalized disc radii presented in Fig.\ref{fig:ratio_curve}-b were determined 
from Eqn.(\ref{visc_orb_ratio}). In this equation, $\frac{f_{visc}}{f_{orb}}$ is taken to be
the normalized break frequency and $\frac{H_d}{R_d}$ is calculated as the thickness of standard 
$\alpha$-disc
\begin{equation}
\frac{H}{R}\approx2.3\times10^{-2}\alpha^{-1/10}R_{10km}^{3/20}M_{1.4}^{-21/40}L_{37}^{3/20}R_{10}^{1/8},
\end{equation}
where  $M_{1.4}$ is the mass of the compact object,$M_1$, in units of $1.4M_\odot$, 
i.e $M_1=M_{1.4}\times1.4M_{\odot}$,
$R_0=10^6R_{10km}$ cm is the radius of the compact object,
 $L_X=GM_1\dot{M}/R_0=10^{37}L_{37}$ erg/s is the X-ray luminosity of the source,
 and $R_d=10^{10}R_{10}$ cm is the outer disc radius. 
 Since the derivation of this formula relies on the thin disc approximation,
 our calculations provide lower limits to the disc sizes. However, the accretion discs of LMXBs
 show many signs of much thicker accretion discs, eg. by possessing additional coronal flows 
covering the thin accretion disc etc., (see \citet{gil05} for details). 
The accretion discs of HMXBs should have similar structures that make the disc seem thicker
because of the nature of accretion discs. However, we do not expect the irradiation 
of the disc in high mass X-ray binary systems to be as effective as in those of low mass X-ray 
binaries; hence, coronal flow would not extend to as large scales as that for $H/R\sim 0.1$ 
which the case for LMXBs \citep{gil05}. Nevertheless, the thickening of the accretion disc 
would not prevent the Roche-lobe overflow systems and the wind accreting
systems being in distinct groups in terms of the low frequency break frequencies in their power
density spectra. The subject of \citet{gil05} is the separation of the low mass 
X-ray binaries into two groups. This is not directly related to our results although we used
a similar method to them, because LMXBs are all Roche-lobe overflow systems. The grouping 
they discovered was related to the orbital period of the LMXBs. Their conclusion 
is that this dichotomy is due to the excitation of 3:1 Lindblad resonances in the wide systems,
which may cause unusual changes in the accretion disc, such as the truncation of the disc
at a specific radius, but does not apply to HMXBs.

\section{Summary}
 
We have estimated the power density estimates using the cosine transform of autocorrelation function, 
a method that can be used for unevenly sampled time series. Using the normalized break frequencies  of 
the sources, we have distinguished two groups of high mass X-ray binaries. 
Those with lower normalized break 
frequencies are more likely to have Roche-lobe-overflow accretion discs. These sources are 
SS 433, OAO 1657-415, SMC X-1, and LMC X-1. On the other hand, 
 Vela X-1, GX 301-2, and 4U 1700-377 have higher normalized break frequencies. These 
sources are wind accretors. We have discussed the compatibility of the observable properties of the sources 
and normalized break frequencies. 
We have deduced that the value of the normalized break frequency is a 
good estimator for distinguishing wind accretors from Roche lobe accretors. This method is
especially useful for classifying black hole binaries according to their accretion mechanisms,
since they cannot be shown on the Corbet diagram where the spin period of the 
compact object is compared with the orbital period of the system.

%\begin{enumerate}
%\end{enumerate}

\begin{acknowledgements}
We would like to thank Marat Gilfanov for useful discussions which helped to improve the paper. 
We appreciate the useful comments and suggestions of the anonymous referee.
We also thank ASTRONS (Astrophysics of Neutron Stars, an FP6 Transfer of Knowledge Project,MTKD-CT-2006-042722) 
for their support throughout our research. AB thanks TUBITAK 1001 project:109T748.
\end{acknowledgements}

\bibliographystyle{aa}
\bibliography{reference_file}

\end{document}